\documentclass[journal=jpclcd,manuscript=letter]{achemso}
\usepackage{amsmath,amssymb}
\usepackage{float}
\usepackage[version=3]{mhchem} 
\usepackage{color} 

\author{Preeti Bhumla$^*$} 
\author{Manish Kumar}
\author{Saswata Bhattacharya}
\email{Preeti.Bhumla@physics.iitd.ac.in[PB], saswata@physics.iitd.ac.in [SB]}
\affiliation{Department of Physics, Indian Institute of Technology Delhi, New Delhi 110016}
\phone{+91-11-2659 1359}
\fax{+91-11-2658 2037}

\title[An \textsf{achemso} demo]
{Theoretical Insights into C$-$H Bond Activation of Methane by Transition Metal Clusters: The Role of Anharmonic Effects}

\begin{document}






\begin{abstract}
Aiming towards materials design for methane activation, we study temperature ($T$), pressure ($p$) dependence of the composition, structure, and stability of metal oxide clusters in a reactive atmosphere using a prototypical model catalyst having wide applications: free transition metal (Ni) clusters in a combined oxygen and methane atmosphere. A robust methodological approach is employed, 
to show that the conventional harmonic approximation miserably fails for this class of materials and capturing anharmonic effects to the vibration free energy contribution is indispensable. To incorporate the anharmonicity in the vibrational free energy, we evaluate the excess free energy of the clusters numerically by thermodynamic integration method with hybrid density functional theory and {\em ab initio} molecular dynamics simulation inputs. We find that the anharmonic effect has a significant impact in detecting the activation of C$-$H bond, whereas the harmonic infrared spectrum completely fails due to the wrong prediction of the vibrational modes.

\begin{tocentry}
	\begin{figure}[H]
		\includegraphics[width=1.0\columnwidth,clip]{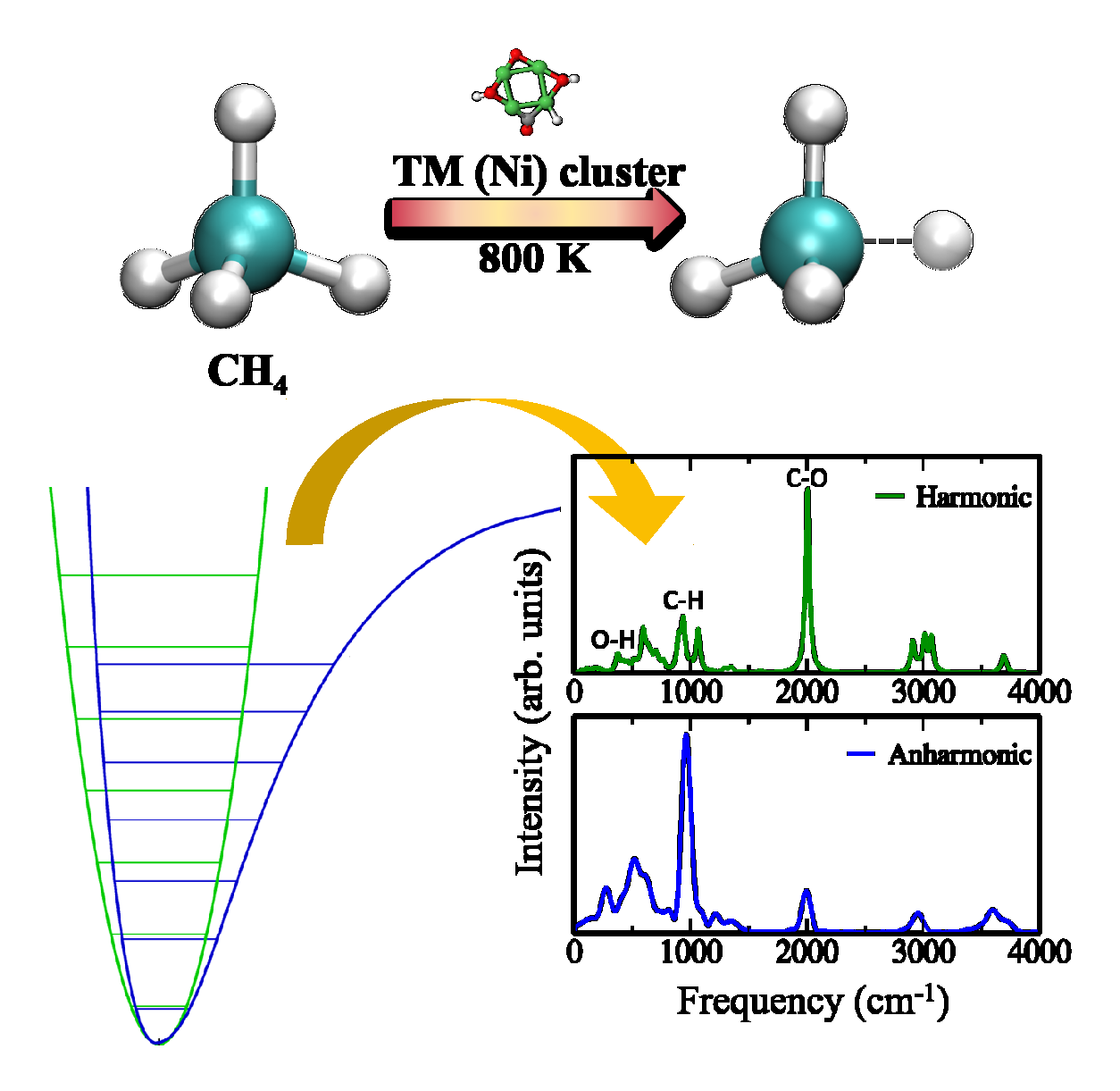}
	\end{figure}
\end{tocentry}
\end{abstract}
Methane is the primary component of natural gas, which is one of the simplest, nearly ubiquitous, low-cost, clean and easily extractable energy sources found in nature.~\cite{doi:10.1021/jacs.8b13552, doi:10.1021/acsanm.8b01256, doi:10.1021/acs.jpcc.5b01724, enger2008review, leenders2015transition} Additionally, methane is also a prominent greenhouse gas. Therefore, it is highly desirable to convert methane into valuable products.~\cite{aasberg2001technologies} Synthesis gas (syngas, a mixture of CO and H$_2$) production from methane is an important route for the effective utilization of abundant natural gas in producing methanol, liquid hydrocarbons, ammonia and dimethyl ether. The efficient activation of methane has been a major challenge as C$-$H bonds in methane possess high bond strength (4.5 eV), low polarizability and negligible electron affinity making it a least reactive hydrocarbon.~\cite{doi:10.1021/acs.accounts.8b00403, doi:10.1021/acs.jpclett.6b02568, doi:10.1021/acs.jpclett.5b00937} Since methane is extremely inert, its conversion to chemical products is difficult. To circumvent this problem, a suitable catalyst must be developed. The catalytic conversion of methane is one of the most appealing fields of study in both academia and industry.~\cite{doi:10.1021/ef0602729, doi:10.1021/acs.iecr.7b00986, C4RA06437B, Hickman343, NEUMANN2004565, ashcroft1990selective, vernon1990partial, doi:10.1021/ie070084l,doi:10.1021/ie980027f, BHARADWAJ1995109, doi:10.1021/jz300069s, doi:10.1021/jz500728d} 

Transition metal (TM) clusters are well known for their efficient homogeneous and heterogeneous catalytic activity.~\cite{aiken1999review, hill1995homogeneous, SUSSFINK199341, doi:10.1021/acs.jpclett.0c00548} This is primarily ascribed to the presence of partially occupied d-shells, which assist in exhibiting multiple oxidation states in their complexes.~\cite{doi:10.1021/acscatal.7b01913,doi:10.1002/adfm.201200615, doi:10.1021/ja044389s} 
In order to understand the activity of a heterogeneous catalyst, one of the most important aspects is to identify the active species and to determine the structure of the catalyst.~\cite{saini2019unraveling} Under reaction conditions, the catalyst comprises of a wide range of structures including the different number of atoms with various oxidation states, all of which could be active to some extent in the catalytic reaction. Therefore, a comprehensive understanding of the catalytic process entails a robust methodological approach that integrates various levels of theory combined into one multi-scale simulation.~\cite{andersen2019practical} 

Numerous experimental and theoretical studies have established that the reactivity of small metal clusters in gas phase varies with the number of atoms.~\cite{doi:10.1021/acs.jpclett.5b01435, doi:10.1021/ja00095a051, PhysRevB.91.241115} It has been found that reducing particle size in the cluster reveals the possibility of several interesting size effects.~\cite{C5CP04281J, doi:10.1021/ja044389s, B502142C, doi:10.1021/jz3018286}                                                               Moreover, the properties of a material change substantially under operational environment, particularly in the atmosphere of reactive molecules. Accordingly, some inevitable questions arise naturally, for e.g., ``which are the species present in the real catalyst and what are their structures?'', and ``how those catalysts change their structure and catalytic properties upon adsorption of different ligand molecules?''. In light of this, there is a justified need to provide theoretical guidance to experiments on the stoichiometry and stability of the clusters under realistic conditions. To better understand the situation theoretically, we consider a prototypical model system of nickel ($\textrm{Ni}_{4}$, as tetrahedron nickel cluster has high selectivity~\cite{doi:10.1021/acs.jpcc.5b01738}) in a reactive atmosphere of $\textrm{O}_2$ and $\textrm{CH}_4$ gas molecules under realistic conditions. Note that Ni-based catalysts, owing to their low cost, high selectivity and high activity, have been extensively employed in catalysis over the past.~\cite{tasker2014recent, keim1990nickel, doi:10.1021/jz5007675} Typically, in the presence of a reactive atmosphere, clusters adsorb surrounding gas molecules and form intermediate phases [Ni$_4$O$_x$(CH$_4$)$_y$] at thermodynamic equilibrium. The latter generally proves to be an active material for various applications in the field of heterogeneous catalysis, making it crucial to have the idea of stable stoichiometries.

In this letter, we have investigated the role of environment [i.e., temperature ($\it{T}$), partial pressure of oxygen ($p_{\textrm{O}_{2}})$ and partial pressure of methane ($p_{\textrm{CH}_{4}}$)] to understand the thermodynamic stability of different configurations of Ni$_4$O$_x$(CH$_4$)$_y$ (0$\leq$$\it{x}$$\leq$8, 0$\leq$$\it{y}$$\leq$3) clusters in a reactive atmosphere of O$_2$ and CH$_4$ molecules. As a first step, a systematic scanning of potential energy surface (PES) is done via cascade genetic algorithm (cGA)~\cite{bhattacharya2014efficient, bhattacharya2013stability, bhattacharya2015computational} approach to obtain the global minimum (GM) configurations of Ni$_4$O$_x$(CH$_4$)$_y$ clusters. Subsequently, we have employed \textit{ab initio} atomistic thermodynamics ($\it{ai}$AT)~\cite{rogal2006ab, doi:10.1021/acs.jpcc.8b08687} in the framework of density functional theory (DFT)~\cite{PhysRev.136.B864, PhysRev.140.A1133} to determine the thermodynamic stability of those configurations under operational conditions. 
To incorporate the anharmonicity in the vibrational free energy contribution to the configurational entropy, we evaluate the excess free energy of the clusters numerically by thermodynamic integration method with {\em ab initio} molecular dynamics (aiMD) simulation inputs. 
On analyzing a large dataset, we show that the conventional harmonic approximation miserably fails to estimate the accurate thermodynamic stability. Therefore, consideration of anharmonic effects is of paramount importance to avoid all the possibilities of missing the stable phases of the clusters. If the anharmonic effects are not included, the stable phases would be destabilized erroneously resulting in inaccurate prediction of the stable phases. Further, we have computed the Infrared (IR) spectra of these stable configurations, which also confirm the anharmonicity in such structures. Besides, the latter has significance in the activation of C$-$H bond, while the harmonic IR spectrum fails to capture it. The sharp peak corresponding to the C$-$H stretching mode (of the activated C$-$H bond) in the anharmonic IR spectrum signifies enhanced dipolar interaction in the C$-$H bond, which results from the localization of charge in C and H atoms of the Ni$_4$O$_7$(CH$_4$)$_2$ cluster, is well captured by the anharmonic IR spectrum. Therefore, to develop a suitable catalyst (with active sites), incorporation of the anharmonic effects is essential in these class of materials. 

After obtaining all low energy isomers corresponding to different configurations of $\textrm{Ni}_{4}\textrm{O}_{x}(\textrm{CH}_{4})_{y}$ clusters from cGA, we study their thermodynamic stability under realistic conditions using \textit{ai}AT approach.
Here, we assume that there is an exchange of atoms between the system (Ni$_4$ cluster) and the surroundings (consisting of $\textrm{O}_2$ and $\textrm{CH}_4$ gas molecules) at finite temperatures and pressures, via the following reaction:
\begin{equation}
\textrm{Ni}_4 + \frac{x}{2}\textrm{O}_2+ y \:\textrm{CH}_4\rightleftharpoons \textrm{Ni}_{4}\textrm{O}_{x}(\textrm{CH}_{4})_{y}
\end{equation}
The Gibbs free energy of formation ($\Delta{G}$) of all the  Ni$_4$O$_x$(CH$_4$)$_y$  structures is then evaluated as a function of $\it{T}$, $p_{\textrm{O}_{2}}$ and $p_{\textrm{CH}_{4}}$ by applying \textit{ai}AT. The composition (for a particular value of $\it{x}$, $\it{y}$) having the minimum Gibbs free energy of formation is most likely to be found in the experiments at a specific $\it{T}$, $p_{\textrm{O}_{2}}$ and $p_{\textrm{CH}_{4}}$. $\Delta{G} (T,p)$ is, therefore, calculated as per the following equation:
\begin{equation}
\Delta{G}(T, p_{\textrm{O}_{2}}, p_{\textrm{CH}_4})=F_{\textrm{Ni}_{4}\textrm{O}_{x}(\textrm{CH}_{4})_{y}} (T)-F_{\textrm{Ni}_4} (T)-x\times\mu_\textrm{O}(T, p_{\textrm{O}_{2}})-y\times\mu_{\textrm{CH}_{4}}(T, p_{\textrm{CH}_{4}})
\label{eqn2}
\end{equation}
Here, $\it{F}$$_{\textrm{Ni}_{4}\textrm{O}_{x}(\textrm{CH}_{4})_{y}} (T)$ and $\it{F}$$_{\textrm{Ni}_4} (T)$ are the Helmholtz free energies of the cluster+ligands [Ni$_4$O$_x$(CH$_4$)$_y$] and the pristine 
[${\textrm{Ni}_4}$] cluster, respectively. $\it{x}$ and $\it{y}$ represent the number of oxygen atoms and methane molecules, that are exchanged with the environment in the reactive atmosphere, respectively. $\mu_\textrm{O}(T, p_{\textrm{O}_{2}})$ and $\mu_{\textrm{CH}_{4}}(T, p_{\textrm{CH}_{4}})$ represent the chemical potential of an oxygen atom ($\mu_\textrm{O}=\frac{1}{2}\mu_{\textrm{O}_{2}}$) and the methane molecule, respectively. The relation of $\mu_\textrm{O}(T, p_{\textrm{O}_{2}})$ with $\it{T}$ and $p_{\textrm{O}_{2}}$ is governed by the ideal (diatomic) gas approximation. The expression is as follows~\cite{basera2019stability}:
\begin{equation}
\begin{aligned}
\begin{split}
\mu_{\textrm{O}_2}\left(T,p_{\textrm{O}_2}\right) & = -{k}_\textrm{B}T \:\textrm{ln} \left[\left(\frac{2\pi \textrm{m}}{{h}^2}\right)^\frac{3}{2} ({k}_\textrm{B}T)^\frac{5}{2}\right]\\& + {k}_\textrm{B}T \:\textrm{ln}\: p_{\textrm{O}_2} - {k}_\textrm{B}T \:\textrm{ln} \left(\frac{8 \pi^2 \textrm{I}_\textrm{A}{k}_\textrm{B}T}{{h}^2} \right)\\& + \frac{{h}\nu_\textrm{OO}}{2} + {k}_\textrm{B}T \:\textrm{ln}\left[ 1 - \exp \left(-\frac{{h}\nu_\textrm{OO}}{{k}_\textrm{B}T}\right)\right]\\& + {{E}^\textrm{DFT}}(\textrm{O}_2) - {k}_\textrm{B}T \:\textrm{ln} \; \mathcal{M} 
+{k}_\textrm{B}T\: \textrm{ln}\, \sigma 
\end{split}
\end{aligned}
\end{equation}
For CH$_4$ molecule, I$_A$=I$_B$=I$_C$=I, and therefore,
\begin{equation}
\begin{aligned}
\begin{split}
\mu_{\textrm{CH}_4}\left(T,p_{\textrm{CH}_4}\right) & = -{k}_\textrm{B}T \:\textrm{ln} \left[\left(\frac{2\pi \textrm{m}}{{h}^2}\right)^\frac{3}{2} ({k}_\textrm{B}T)^\frac{5}{2}\right]\\&  + {k}_\textrm{B}T \:\textrm{ln}\: p_{\textrm{CH}_4} - {k}_\textrm{B}T \:\textrm{ln} \left(\frac{8 \pi^2 \textrm{I}{k}_\textrm{B}T}{{h}^2} \right)^\frac{3}{2} \\& + \frac{h\nu_{{\textrm{CH}}}}{2} + {k}_\textrm{B}T \:\textrm{ln}\left[ 1 - \exp \left(-\frac{{h}\nu_{{\textrm{CH}}}}{{k}_\textrm{B}T}\right)\right] \\& +{{E}^\textrm{DFT}}(\textrm{CH}_4)- {k}_\textrm{B}T \:\textrm{ln} \; \mathcal{M} 
+{k}_\textrm{B}T\: \textrm{ln} \; \sigma        
\end{split}
\end{aligned}
\end{equation}

\noindent Here $k_{\textrm{B}}$, $\it{h}$, ${{E}^\textrm{DFT}}$, $\nu_{OO}$ and $\nu_{{CH}}$ are respectively the Boltzmann constant, Planck constant, total DFT energy and stretching frequencies of O--O and C--H bonds. m, I, $\mathcal{M}$ and $\sigma$ represent the mass, moment of inertia, spin multiplicity and symmetry no. of the molecule, respectively.

The Helmholtz free energies $\it{F}$$_{\textrm{Ni}_{4}\textrm{O}_{x}(\textrm{CH}_{4})_{y}} (T)$ and $\it{F}$$_{\textrm{Ni}_4}(T)$ consist of respective total DFT energy along with their free energy contributions from translational, rotational, vibrational, symmetry and spin-degeneracy terms~\cite{rogal2006ab}. It has been noticed that total DFT energy is the dominant term, which is evaluated in its ground state configuration with respect to both geometry and spin state. 
The rest of the terms, except contribution from vibrational degrees of freedom ($F_{vibs}$), are usually considered as invariant since they do not change much (and even if they change, the order is insignificant) due to the dependence on most of the constant terms viz. mass, moment of inertia, universal constants, etc. However, the vibrational contribution is dependent on frequencies of vibration, which are unique for a given structure. Thus, the Helmholtz free energy can be written as follows:
\begin{equation}
F(T)={E}^\textrm{DFT} + F_{vibs} + \Delta
\label{eqn5}
\end{equation}

$\Delta$ is considered to be the constant term. At low temperature, $F_{vibs}$ usually contributes at the first order after the decimal for a small cluster of few atoms. Thus, while computing $\Delta G (T,p)$, since we take differences of two free energy expressions (i.e. a system with ligands and system without ligands), we  assume this to be very small and therefore, can be neglected. However, there exist some systems, where $F_{vibs}$ contributes significantly even after taking the difference of two such terms to compute $\Delta G (T,p)$.~\cite{doi:10.1021/acs.jpcc.8b08687} In view of this, though a significant number of works have neglected $F_{vibs}$, but it is not recommended. Here, we have estimated the role of $F_{vibs}$ via state-of-the-art theoretical techniques at various level of accuracy. 

\begin{figure}[!htp]
	\includegraphics[width=1.0\columnwidth,clip]{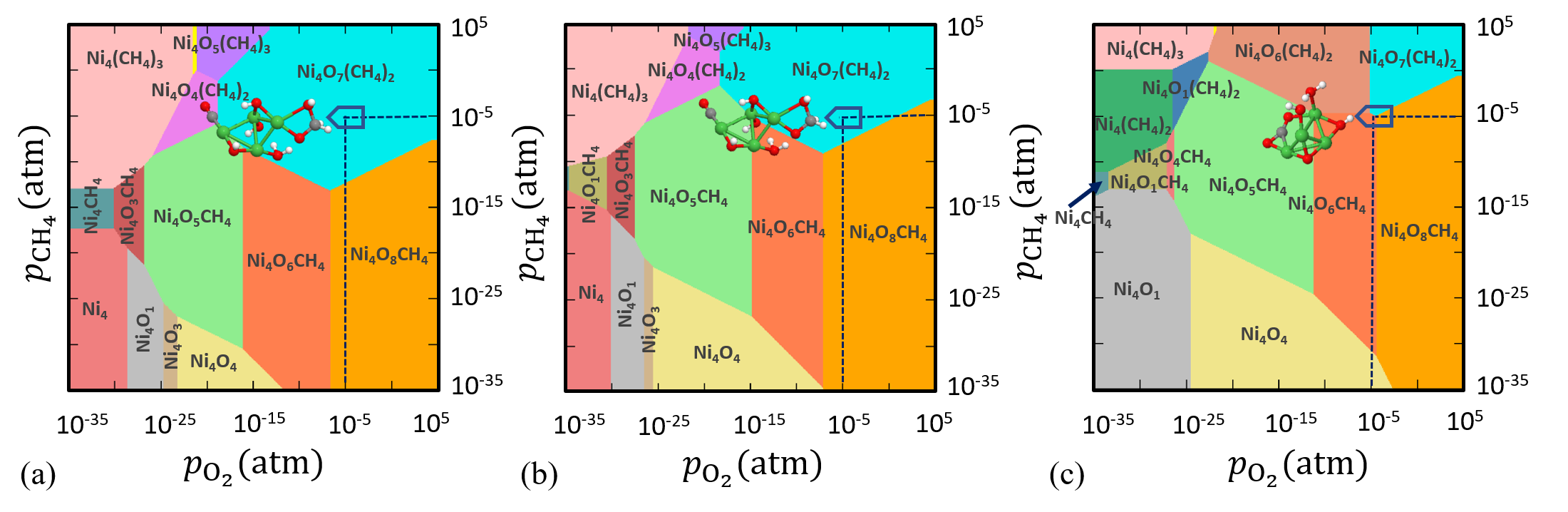}
	\caption{2D projection of 3D phase diagram obtained for Ni$_4$O$_x$(CH$_4$)$_y$ clusters in the reactive atmosphere of O$_2$ and CH$_4$. In this plot $\Delta G(T,p)$ is computed (a) when only DFT total energies are included, (b) DFT+F$_{vibs}^{harmonic}$ are included and (c) DFT+F$_{vibs}^{anharmonic}$ are included to compute $F(T)$ of respective configurations as shown in Equation~\ref{eqn2}. Colored regions show the most stable compositions in a wide range of pressure at $\it{T}$\;=\;800 K.}
	\label{pic}
\end{figure}

Using Equation~\ref{eqn2}, we have obtained the 3D phase diagram ($p_{\textrm{O}_2}$ vs $p_{\textrm{CH}_4}$ vs $\Delta G$(\textit{T},\textit{p})) at an experimentally relevant $T$ (here, 800 K) by taking its 2D projection after aligning negative $\Delta G (T,p)$ axis to be vertically up. We have considered all the configurations of Ni$_4$O$_x$(CH$_4$)$_y$ clusters. Note that only those phases that minimize the $\Delta G (T,p)$ at a specific $p_{\textrm {O}_2}$, $p_{\textrm{CH}_4}$ and $T=800$ K, are visible (see Figure \ref{pic}). Each color in the phase diagram represents a stable configuration of the catalyst. All the phase diagrams are constructed at $\it{T}$\;=\;800\;K as it is a suitable temperature for methane activation.

Herein, we have implemented a suite of three state-of-the-art techniques to plot $\Delta G$(\textit{T},\textit{p}). The first one is without any explicit contribution of $F_{vibs}$ (as in Equation~\ref{eqn5}) i.e., only total DFT energy (\textit{E}$^{\textrm{DFT}}$) of the cluster with and without ligands is considered (see Figure~\ref{pic}a). In the second case, we have duly considered $F_{vibs}$ upto harmonic approximation to calculate $\Delta G$(\textit{T},\textit{p}). $F_{vibs}^{harmonic}$ is computed using the following equation~\cite{bhattacharya2014efficient}:
\begin{equation}
F_{vibs}^{harmonic} = \sum_{i} \frac{h\nu_{i}}{2}+{k}_{\textrm{B}}T~{\textrm{ln}}\left[1-\exp\left(\frac{h\nu_{i}}{k_{\textrm{B}}T}\right)\right]
\end{equation}
Note that after adding $F_{vibs}^{harmonic}$ with \textit{E}$^{\textrm{DFT}}$ (as in Equation~\ref{eqn5}), a new phase is introduced along with slight rearrangement of the existing phases, especially near the boundary region of competing configurations (see Figure~\ref{pic}b). However, despite some small changes in Figure~\ref{pic}a and~\ref{pic}b, we do not see any significant difference to identify the most stable phases at experimentally realistic pressure range. In this region, Ni$_4$O$_6$CH$_4$, Ni$_4$O$_7$(CH$_4$)$_2$ and Ni$_4$O$_8$CH$_4$ are the stable phases and if we see at the region, where $p_{\textrm {O}_2}$ = $p_{\textrm{CH}_4}$ = 10$^{-5}$ atm ($\it{T}$\;=\;800\;K), Ni$_4$O$_7$(CH$_4$)$_2$ comes out to be the most stable phase (see Figure~\ref{pic}a and~\ref{pic}b) irrespective of $F_{vibs}^{harmonic}$ is taken into consideration or not. 
Now here, it is assumed that at $\it{T}$\;=\;800\;K, the oscillations are constrained to vibrate under a harmonic potential. However, the real system does not necessarily follow this assumption. And if so, the real anharmonic potential can be very different from harmonic potential. In such case, the expression for $F_{vibs}$ can vary from one configuration to the other. 

Therefore, in an attempt to refine the expression of $\Delta G (T,p)$ at finite $T, p$, we have included anharmonic effects in the potential energy surface (see Figure \ref{pic}c). In order to quantitatively account for the anharmonic effects, we have performed the thermodynamic integration taking input from {\em aiMD} simulation to evaluate the excess free energy of clusters. Here, we have assumed that at low $T$ (e.g. 10~K), both harmonic and anharmonic potentials do not diverge much. Taking such low $T$ as our reference state, the Helmholtz free energy $F(T)$ is calculated according to the following equation (for detailed derivation, see section I in Supporting Information (SI)):
\begin{equation}
F(T)=\underbrace{E^{\textrm{DFT}} +U^{ZPE}}_{U^{ref}}+\frac{T}{T_\circ}F_{vibs}^{harmonic}(T_\circ)-T\underbrace{\int_{T_\circ}^{T}\frac{dT}{T^2}(\langle U \rangle_{T}-U^{ref})}_{thermodynamic\;\; integration}-k_BT \:\frac{N}{2}\ln\frac{T}{T_\circ}
\end{equation}
where $\it{T}$$_\circ$ and $\it{T}$ represent the initial and final temperatures, respectively. \textit{E}$^{\textrm{DFT}}$, \textit{U}$^{ref}$, $F_{vibs}^{harmonic}(T_\circ)$, N and $\langle U \rangle_{T}$ are respectively the total DFT energy, zero point energy, Helmholtz free energy at temperature $\it{T}$$_\circ$ (10\;K) under harmonic approximation, total number of atoms and canonical average of the total energy at temperature $\it{T}$ (800\;K) of the clusters. We have run $\it{ai}$MD simulations in canonical ensemble for 8 ps at five different temperatures, from $\it{T}$\;=\;10\;K to $\it{T}$\;=\;800\;K to obtain the average energy ($\langle U \rangle_{T}$). After that, we have performed quadratic curve fitting for this data and numerically integrated the corresponding function over the limits, $\it{T}$$_\circ$\;=\;10\;K to $\it{T}$ =\;800\;K to get the value of $F(T)$ at $T=800$~K. After evaluating $F(T)$, we have minimized $\Delta G (T,p)$ using the same aforementioned procedure and obtained the phase diagram with the anharmonic effects. Interestingly, we have noticed, a completely new phase viz. Ni$_4$O$_6$(CH$_4$)$_2$ appears to be stable alongside three existing phases [viz. Ni$_4$O$_6$CH$_4$, Ni$_4$O$_8$CH$_4$ and Ni$_4$O$_7$(CH$_4$)$_2$] at reaction condition ($p_{\textrm{O}_2}$ = $p_{\textrm{CH}_4}$ = 10$^{-5}$ atm and $\it{T}$\;=\;800\;K). On comparing Figure~\ref{pic}a, \ref{pic}b and \ref{pic}c, we infer that stable phases have not only been destabilized erroneously but also the new phases have a high probability of being missed at reaction conditions, if the anharmonic effects are not taken into consideration for this class of materials. Hence, it manifests that the inclusion of anharmonicity in these clusters affects the thermodynamic stability under operational conditions. Next, we have shown two important applications of this finding by computing the IR spectra of two test cases: (i)  Ni$_4$O$_6$(CH$_4$)$_2$ and (ii) Ni$_4$O$_7$(CH$_4$)$_2$.

IR spectroscopy covers the infrared region of the electromagnetic spectrum with frequencies ranging from 4000 cm$^{-1}$ to 40 cm$^{-1}$~\cite{stuart2000infrared, weng2000mechanistic, doi:10.1002/anie.201912668, C4CY00486H, wang2018copper}.
\begin{figure}[!htp]
	\includegraphics[width=0.6\columnwidth,clip]{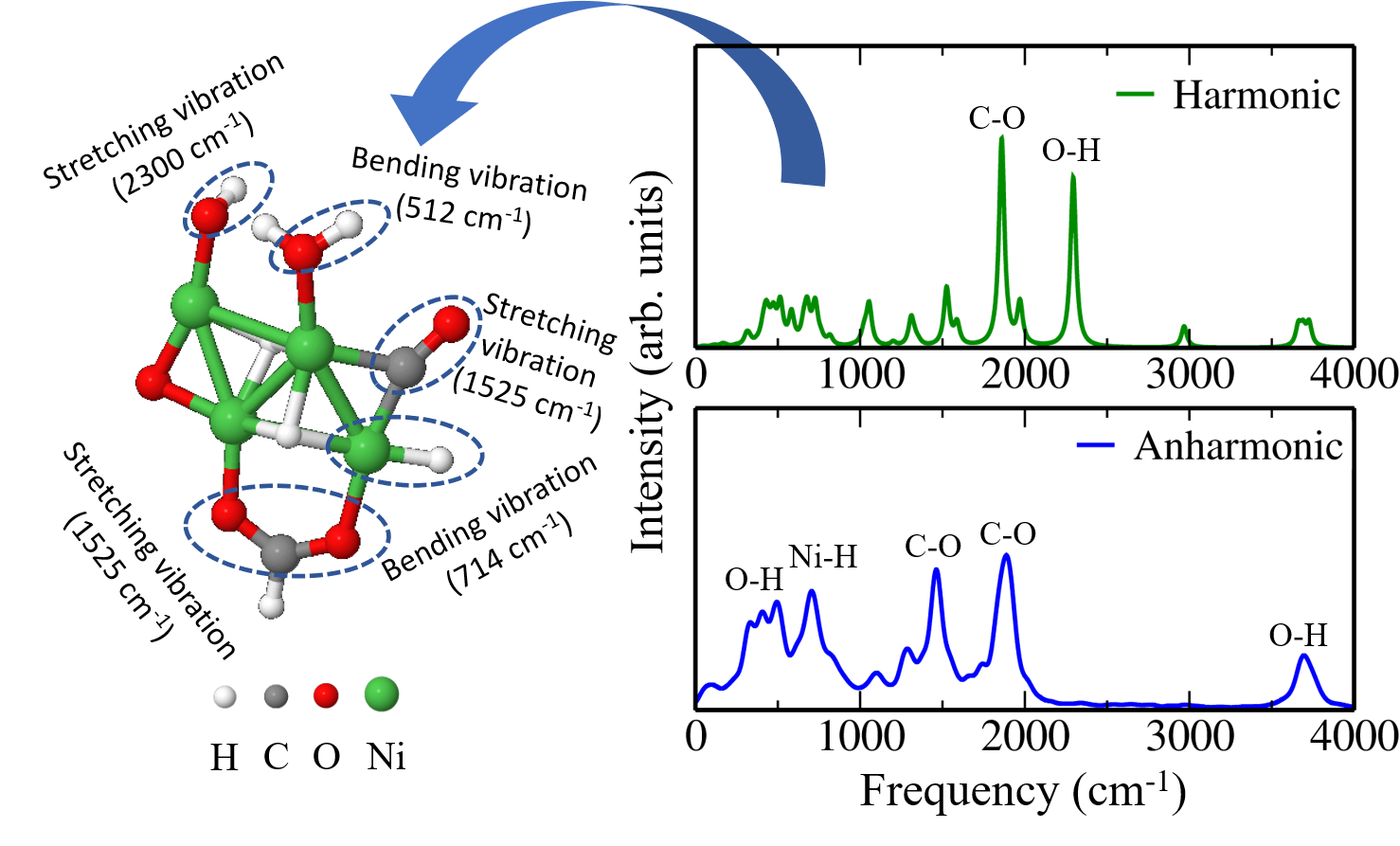}
	\caption{Infrared (IR) spectra of Ni$_4$O$_6$(CH$_4$)$_2$ for both harmonic (upper panel) as well as the anharmonic (lower panel) case. The possible vibrational modes are also shown corresponding to those respective peaks.}
	\label{2}
\end{figure}
In IR spectroscopy, specific frequencies are absorbed by the molecules that are the characteristic of their structure. Here, we have simulated the IR spectra of one of the clusters viz. Ni$_4$O$_6$(CH$_4$)$_2$, that is explicitly stable on including the anharmonic contribution to the free energy, to determine its characteristic vibrational normal modes. For this, we have run 8\;ps \textit{ai}MD simulation in the canonical ensemble with Bussi-Donadio-Parrinello (BDP)~\cite{ruiz2018effect} thermostat. From Figure \ref{2}, we have noticed significant dissimilarities between the harmonic and anharmonic IR spectra of Ni$_4$O$_6$(CH$_4$)$_2$. Aside from the usual difference in peak intensities, the O$-$H stretching mode as per harmonic IR analysis near 2300 cm$^{-1}$ (see Figure~\ref{2} upper panel) is just a negligible hump in anharmonic IR (see Figure~\ref{2} lower panel). In addition to this, there is an intense C$-$O bending mode visible around 1500 cm$^{-1}$, which is not that prominent as per the harmonic analysis. Hence, it is evident that there is a fundamental difference in the characteristic frequencies of vibration of this structure as computed with harmonic approximation and that of after capturing the anharmonic effects. As a result, they contribute differently to the free energy of vibration. This makes Ni$_4$O$_6$(CH$_4$)$_2$ stable in the anharmonic case, whereas unstable under the harmonic approximation. Note that we have taken just a prototypical model system here viz. Ni$_4$ cluster to study its stable phases under the reactive atmosphere of O$_2$ and CH$_4$. Nevertheless, this model system is relevant and sufficient to convey the underlying message that there is a high chance of leaving important stable phases of the catalyst while ignoring the anharmonic effects during reaction condition.  

\begin{figure}[!htp]
	\includegraphics[width=1.0\columnwidth,clip]{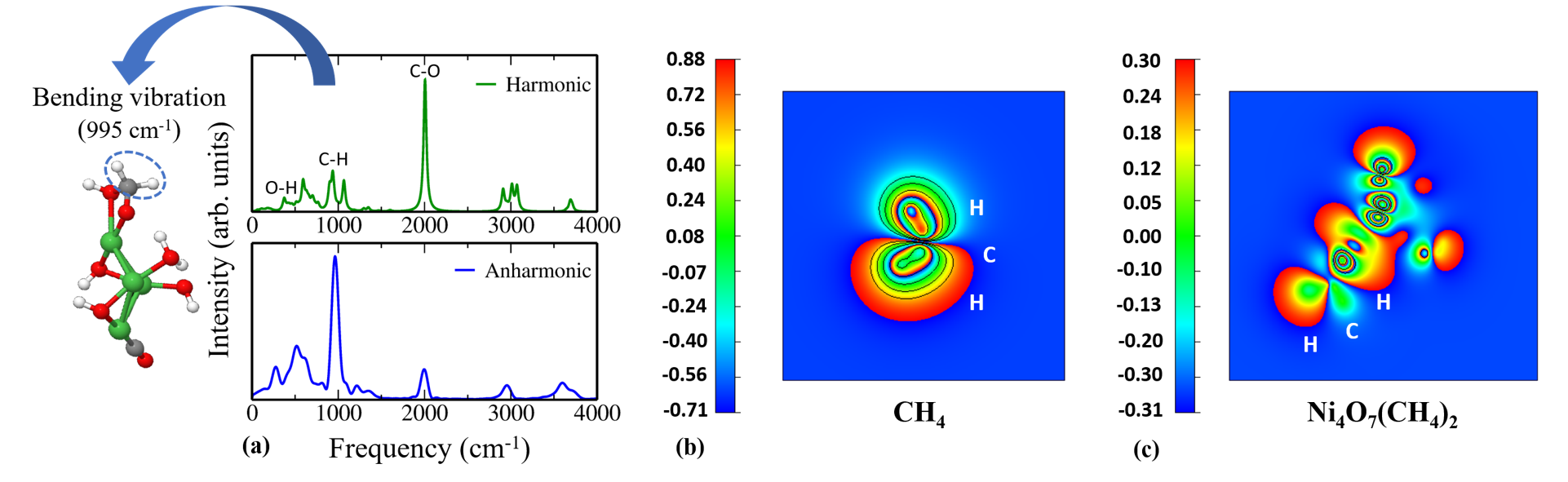}
	\caption{(a) Infrared (IR) spectra of Ni$_4$O$_7$(CH$_4$)$_2$ for both harmonic as well as the anharmonic case. Contour plots of electronic charge density associated with (001) plane of (b) CH$_4$ (delocalization of charge within C$-$H bonds) and (c) Ni$_4$O$_7$(CH$_4$)$_2$ cluster (localization of charge within C$-$H bonds).}
	\label{1}
\end{figure}

Apart from the above important facts, we notice the additional significance of including anharmonic effects related to C$-$H bond activation efficiency. For this, we have considered a test case viz. Ni$_4$O$_7$(CH$_4$)$_2$ cluster, which is stable in all the three cases as shown in Figure~\ref{pic}a, ~\ref{pic}b and ~\ref{pic}c. We have plotted its IR spectra (harmonic vs anharmonic) and compared in Figure~\ref{1}a. From Figure \ref{1}a, we have noticed that O$-$H stretching presents significant anharmonic red-shifts in comparison to harmonic case around 273 cm$^{-1}$. These red-shift corrections lead to change in IR spectrum shape due to a reorganization of the vibrational modes. Primarily, we have observed some remarkable dissimilarities between harmonic and anharmonic IR spectra, for e.g., the intensity of C$-$O stretching peak has reduced significantly after the inclusion of anharmonic effects. Moreover, in the anharmonic IR spectrum, we have found an intense peak around 995 cm$^{-1}$ corresponding to C$-$H bending vibration. This type of highly intense IR absorption is due to the change in dipole moment that occurs during a vibration, especially when the bond is highly polar in nature so that its dipole moment changes considerably as the bond stretches. Therefore, according to the anharmonic IR spectrum, this structure is supposed to be a good catalyst for C$-$H bond activation of methane. However, the harmonic IR spectrum completely fails to capture this information. 

To validate this enhance dipolar interaction into this structure, we have plotted the charge density of Ni$_4$O$_7$(CH$_4$)$_2$ cluster and compared the same with CH$_4$ (see Figure \ref{1}b and \ref{1}c). 
To obtain the charge density contour analysis, we have plotted the electron charge density for CH$_4$ and  Ni$_4$O$_7$(CH$_4$)$_2$ cluster for the electronic levels near respective Highest Occupied Molecular Orbitals (HOMO). The constant slicing plane is  chosen such that both the C and H atoms of the C$-$H bond is covered. The value of electron charge density varies from maximum (red color) to minimum (blue color). Now, if we notice the nature of the C$-$H bond in either case, we can clearly see the difference in charge localization. In conventional CH$_4$, the C$-$H bond is purely covalent, which makes it rather inert to get functionalized easily. However, in the Ni$_4$O$_7$(CH$_4$)$_2$ cluster, the C$-$H bond is very much polar with the localized charge on C and H respectively. This unusual localization of charge in the C$-$H bond gives rise to enhanced dipolar interactions (see Figure \ref{1}c), and as a consequence of this, Ni$_4$ happens to be a reliable catalyst by activating the C$-$H bonds in methane. However, if this entire analysis is done using only harmonic approximation, this stable configuration would not even be considered for C$-$H bond activation as its peak in the IR spectrum is pretty small and delocalized. This further concludes the importance of capturing the anharmonic contribution to this class of materials.
 
In summary, we have carried out state-of-the-art hybrid density functional theory (DFT) calculations combined with $\it{ab}$ ${initio}$ atomistic thermodynamics ($\it{ai}$AT) and $\it{ab}$ $\it{initio}$ molecular dynamics ($\it{ai}$MD) simulation to see how the thermodynamic stability of TM  oxide clusters changes as a function of temperature and pressure ($\it{T}$, $p_{\textrm{O}_{2}}$ and $p_{\textrm{CH}_{4}}$). While finding the accurate thermodynamic stability, we have seen that inclusion of anharmonicity introduces new stable phases that are entirely ignored by the DFT and DFT+F$_{vibs}^{harmonic}$. This has a significant impact in detecting activation of C$-$H bond, where the harmonic IR is unable to capture the correct vibrational modes. The key point, which emerges out of these studies, is that towards understanding the activation of the stable C$-$H bonds in methane using a metal oxide cluster as a catalyst, capturing the anharmonic effects is essential for this class of materials.


\section{Methodology}
We have generated a large data set of Ni$_4$O$_x$(CH$_4$)$_y$ (0$\leq$$\it{x}$$\leq$8, 0$\leq$$\it{y}$$\leq$3) clusters. We have varied the value of $x$ and $y$ ($x=$ no. of oxygen  atoms, $y=$ no. of CH$_4$ molecules) from zero to the saturation value, which means $x$ and $y$ values are increased with all possible combinations until no more O-atom/CH$_4$ molecule can be absorbed by the cluster. As a first step, we have used a massively parallel cascade genetic algorithm (cGA) to thoroughly scan the potential energy surface (PES) in determining all possible low-energy structures (including the global minimum). The term ``cascade'' means a multi-stepped algorithm, where successive steps employ higher level of theory and each of the next level takes information obtained at its immediate lower level. Typically, a cGA algorithm starts with classical force field and goes upto density functional theory (DFT) with hybrid exchange and correlation ($\epsilon_{xc}$) functionals. Note that it is reported that PBE $\epsilon_{xc}$ functional ~\cite{PhysRevLett.77.3865} highly overestimates stability of clusters containing larger concentration of O-atoms~\cite{bhattacharya2013stability, PhysRevB.91.241115, saini2018structure, saini2019unraveling}. This results in a qualitatively wrong prediction of O$_2$ adsorption for O-rich cases. Such behaviour is not confirmed by more advanced hybrid $\epsilon_{xc}$ functionals [e.g HSE06~\cite{heyd2003hybrid}, PBE0~\cite{perdew1996rationale}] as employed in our calculations. Moreover, the spin states of the clusters are also different as found by PBE and PBE0/HSE06. In view of this, in the cascade algorithm, we have only optimized with PBE but the energetics are computed with PBE0~\cite{perdew1996rationale} to evaluate the fitness function of the cluster. For details of this cGA implementation, accuracy and validation, we recommend our previous studies as given in Ref.~\cite{bhattacharya2013stability,bhattacharya2014efficient,bhattacharya2015computational}. 

All DFT calculations have been performed using FHI-aims code, employing an all-electron code with numerical atom centered basis sets.~\cite{BLUM20092175} Considering the fact that first-principles based calculations are computationally demanding, lighter (viz. light settings with tier 2 basis set~\cite{BLUM20092175}) DFT settings have been implemented in the cGA to find the global minimum structures. The atomic zero-order regular approximation (ZORA) is used for the scalar relativistic correction.~\cite{doi:10.1063/1.466059} The vdW correction is calculated according to the Tkatchenko-Scheffler scheme.~\cite{tkatchenko2009accurate} The low energy structures obtained from the cGA are further optimized with PBE at higher level settings (viz. tight settings with tier 2 basis set~\cite{BLUM20092175}). The atomic forces are converged up to 10$^{-5}$ eV/${\textrm \AA}$. Finally, the total single point energy is calculated on top of this optimized structure using PBE0 hybrid $\epsilon_{xc}$ functional (see further details for validation of $\epsilon_{xc}$ functionals as in section II in SI). The vibrational frequencies are determined of the stable compositions under harmonic approximation using finite displacement method. 

Next, to capture the anharmonic effects using thermodynamic integration method, we have carried out $\it{ab}$ $\it{initio}$ molecular dynamics (MD) simulations for 8 ps each at different temperatures namely $\it{T}$\; = 50\;K, 100\;K, 300\;K, 600\;K and 800\;K in canonical ensemble (i.e., one with constant temperature and volume). We have employed Velocity Verlet scheme~\cite{PhysRevE.59.3733} for the integration of Newtonian equations with a time-step of 1\;fs and the temperature of the system is controlled using Nose-Hoover thermostat.~\cite{evans1985nose} (for details see section III in SI)
 
\section{Acknowledgement}
PB acknowledges UGC, India, for the junior research fellowship [1392/(CSIR-UGC NET JUNE 2018)]. MK acknowledges CSIR, India, for the senior research fellowship [grant no. 09/086(1292)/2017-EMR-I]. SB acknowledges the financial support from SERB under core research grant (grant no. CRG/2019/000647). PB acknowledges Shikha Saini for helpful discussions. We acknowledge the High Performance Computing (HPC) facility at IIT Delhi for computational resources.

\section{Supporting Information Available}
Details of choice of functionals, temperature control by Nose-Hoover thermostat and thermodynamic integration can be found in the supporting information file.

\bibliography{references2}


\end{document}


\begin{center}
{\Large \bf Supplemental Material}\\ 
\end{center}
\begin{enumerate}[\bf I.]
	\item Thermodynamic integration for Helmholtz free energy evaluation 
	\item Comparison of functionals (PBE and HSE06) in finding global minimum structures
	\item Temperature control using Nose-Hoover thermostat

\end{enumerate}
\vspace*{12pt}
\clearpage

\section{Thermodynamic integration for Helmholtz free energy evaluation}
\textrm{We start from the fundamental relation between Helmholtz free energy \textit{F}(\textit{T}) and partition function \textit{Z}}.
\begin{equation}
{F}(T)=-{k}_\textrm{B}T \ln {Z} =-{k}_\textrm{B}T \ln \int e^{-\beta H} {dp\:dr}
\end{equation}
where $\it{H}$ (i.e., kinetic energy, \textit{K}, plus potential energy, \textit{U}) is the Hamiltonian of the system (cluster), and $\beta=1/k_BT$. This on integration gives us
\begin{equation}
F(T)=-k_BT \ln \int e^{-\beta U} dr + Nk_BT \ln \Lambda(T)
\end{equation}
where \textit{N} is number of degrees of freedom and $\Lambda(T)$
is the result of integration over momentum space and always valid only if the system is classical. In terms of $\beta$, the expression for $\Lambda(\beta)$ is: $\Lambda(\beta)= \sqrt{\frac{h^2 \beta}{2\pi m}}$ \\
Equation 1 can then be written as:
\begin{equation}
\beta F(\beta)= - \ln \int e^{-\beta U} dr + N \ln \Lambda(\beta)
\end{equation}
\begin{equation}
\frac{\partial[\beta F (\beta)]}{\partial\beta}=\frac{\frac{\partial}{\partial\beta} \int e^{-\beta U} dr } {\int e^{-\beta U} dr} + N \frac{\frac{\partial}{\partial \beta}\Lambda(\beta)}{\Lambda(\beta)} = \frac{\frac{\partial}{\partial \beta} \int e^{-\beta U} dr}{Z}+\frac{N}{2\beta} = \langle U \rangle_{\beta} +\frac{N}{2\beta}
\end{equation}

\begin{equation}
\beta F(\beta)= \beta_{\circ} F(\beta_{\circ}) + \int_{\beta_{\circ}}^{\beta} \langle U \rangle_{\beta}\: d\beta + \frac{N}{2}\int_{\beta_{\circ}}^{\beta}\frac{\partial\beta}{\beta}
\end{equation}
Now \textit{F($\beta_{\circ}$)} can be written as:
\begin{equation}
F(\beta_{\circ})=\underbrace{E^{\textrm{DFT}} + U^{ZPE}}_{U^{ref}} + F^{vib}({\beta_\circ})
\end{equation}
where from quantum approach, we have
\begin{equation}
F^{vib}_{quantum}(\beta)=\beta^{-1}\sum_{i=1}^{N}\ln(1-e^{-\beta h\nu_{i}})
\end{equation}
and from classical approach, we have
\begin{equation}
F^{vib}_{classic}(\beta)=\beta^{-1}\sum_{i=1}^{N}\ln(\beta h\nu_{i})
\end{equation}
Therefore, we get the final expression of $\beta$F$(\beta)$ as shown below:
\begin{equation}
\beta F(\beta)= \beta_{\circ} U^{ref} +\beta_{\circ} F^{vib}(\beta_{\circ}) + \int_{\beta_{\circ}}^{\beta} \langle U \rangle_{\beta}\: d\beta + \frac{N}{2}\ln \frac{\beta}{\beta_{\circ}}
\end{equation}
We call the above equation as our master equation and we will approximate this equation as given below. \\
Here, we write $\langle$U$\rangle_{\beta}$ as follows:
\begin{equation}
\langle U \rangle_{\beta} \rightarrow (\langle U \rangle_{\beta}-U^{ref})
\end{equation}
Therefore, on substituting this into Equation 9, we get
\begin{equation}
\beta F (\beta) =\beta_{\circ} U^{ref} + \beta_{\circ} F^{vib}(\beta_{\circ}) + (\beta-\beta_{\circ}) U^{ref} + \int_{\beta_{\circ}}^{\beta}(\langle U \rangle_{\beta}-U^{ref})\: d\beta+ \frac{N}{2} \ln \frac{\beta}{\beta_{\circ}}
\end{equation}
\begin{equation}
=\beta U^{ref} + \beta_{\circ} F^{vib}(\beta_{\circ}) + \int_{\beta_{\circ}}^{\beta}(\langle U \rangle_{\beta}-U^{ref})\: d\beta + \frac{N}{2} \ln \frac{\beta}{\beta_{\circ}}
\end{equation}
This can be rewritten in terms of \textit{T}=1$/k_B\beta$ as follows:
\begin{equation}
F(T)=E^{\textrm{DFT}} + U ^{ZPE} +\frac{T}{T_\circ}F^{vib}(T_\circ) - T\int_{T_\circ}^{T}\frac{dT}{T^2}(\langle U \rangle_{T}-U^{ref})-k_BT \:\:\frac{N}{2}\ln\frac{T}{T_\circ}
\end{equation}

\section{Comparison of functionals (PBE and HSE06) in finding global minimum structures}

\begin{figure}[H]
	\centering
	\includegraphics[width=0.65\textwidth]{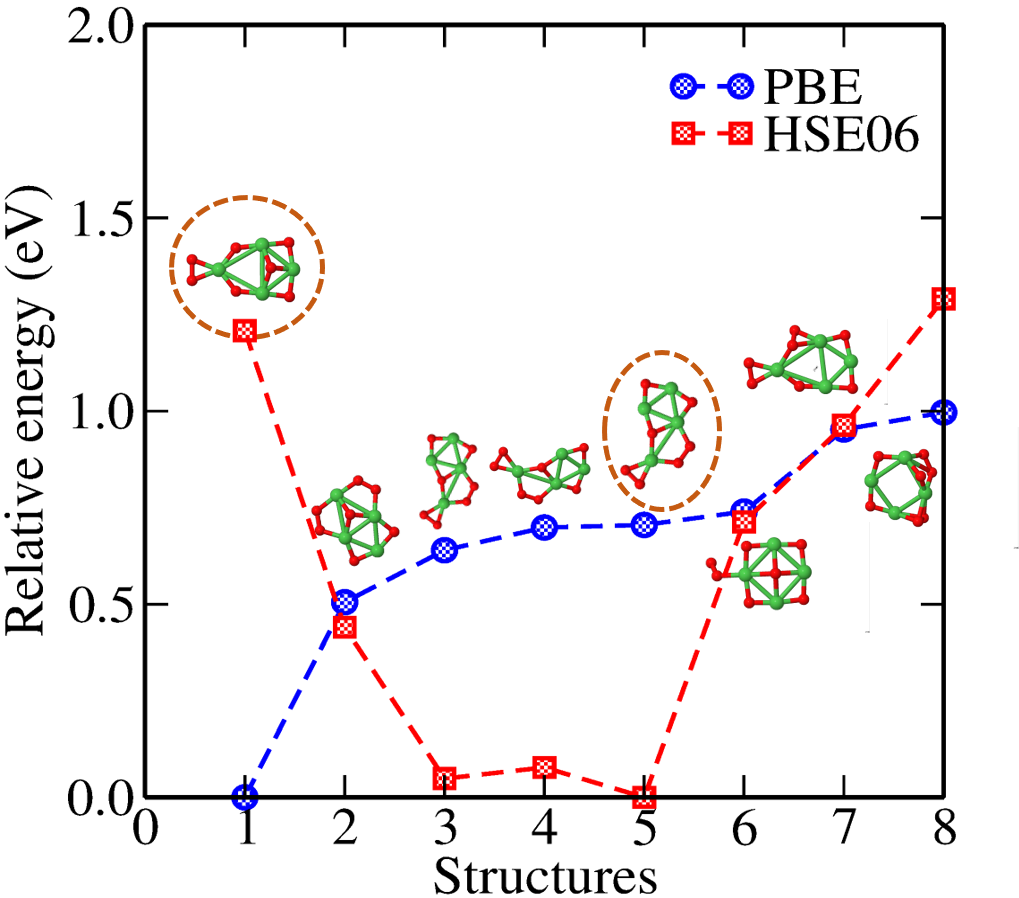}
	\caption{Structures of different isomers of Ni$_4$O$_7$ clusters obtained from PBE (represented by dashed red line) and HSE06 (represented by dashed blue line) exchange-correlation ($\epsilon_{xc}$) functionals. Dashed circles represent the global minima from the two functionals.}
	\label{energy}
\end{figure}

\begin{figure}[H]
	\centering
	\includegraphics[width=0.65\textwidth]{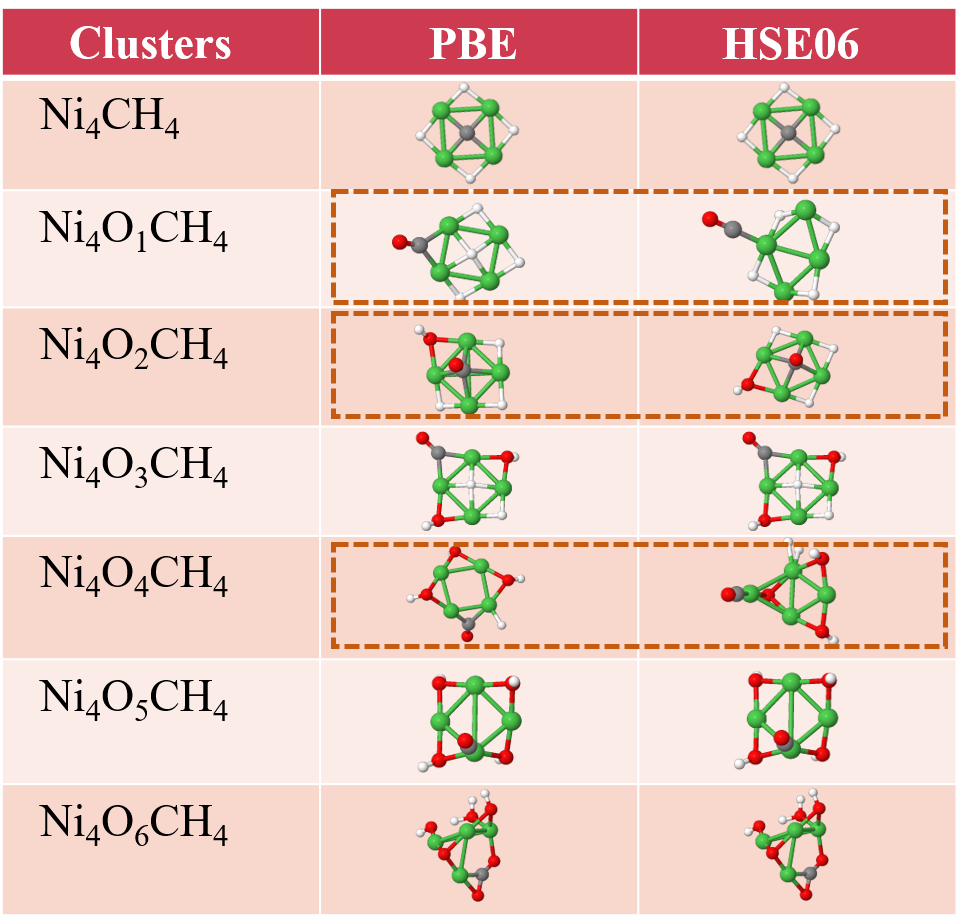}
	\caption{Global minimum structures of Ni$_4$O$_x$CH$_4$ clusters obtained from PBE and HSE06 $\epsilon_{xc}$ functionals. Dashed rectangle indicates the clusters having different global minima.}
	\label{table}
\end{figure}

 We have determined the energy of different isomers of Ni$_4$O$_7$ clusters from both the $\epsilon_{xc}$ functionals viz. PBE~\cite{PhysRevLett.77.3865} and HSE06~\cite{heyd2003hybrid}. From Figure \ref{energy}, we have found that PBE $\epsilon_{xc}$ functional predicts structure 1 as the global minimum isomer while structure 5 is the global minimum isomer obtained from HSE06 $\epsilon_{xc}$ functional.
 Further, we have calculated the energy of Ni$_4$O$_x$CH$_4$ (0$\leq$$\it{x}$$\leq$6) series of clusters from PBE as well as HSE06 $\epsilon_{xc}$ functionals (see Figure \ref{table}). We have noticed different trends of energy from both functionals. For Ni$_4$O$_1$CH$_4$, Ni$_4$O$_2$CH$_4$ and Ni$_4$O$_4$CH$_4$ clusters, PBE and HSE06 $\epsilon_{xc}$ functionals predict different global minimum structures. As it is already well established that for oxygen molecule, HSE06 being advanced functional gives correct binding energy~\cite{bhattacharya2013stability}, therefore, we have chosen HSE06 for our oxide based clusters.

\section{Temperature control using Nose-Hoover thermostat}
\begin{figure}[H]
	\centering
	\includegraphics[width=0.65\textwidth]{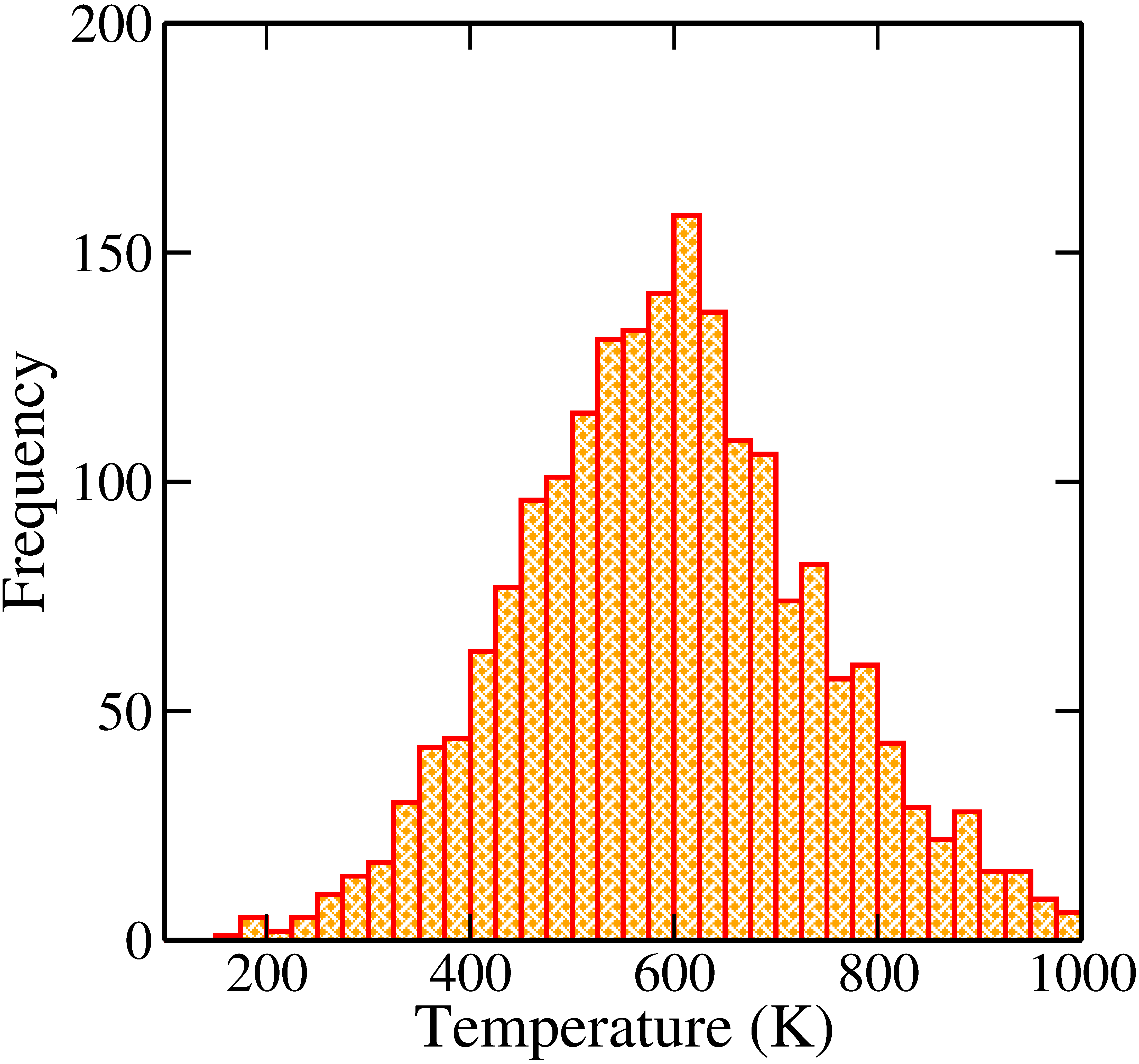}
	\caption{Histogram of $\it{ai}$MD simulation for Ni$_4$O$_6$CH$_4$ cluster at $\it{T}$\;=\;600\;K.}
	\label{his}
\end{figure}
Here, we represent a test case of Ni$_4$O$_x$(CH$_4$)$_y$ cluster to validate our \textit{ab initio} molecular dynamics ($\it{ai}$MD) simulations. We have carried out $\it{ai}$MD simulations at five different temperatures in canonical ensemble with time and time-step being 8 ps and 1 fs, respectively. Temperature control is realized by employing the Nose-Hoover thermostat~\cite{evans1985nose}. Figure \ref{his} shows the histogram that we have obtained from $\it{ai}$MD simulation for Ni$_4$O$_6$CH$_4$ cluster at $\it{T}$\;=\;600\;K. From the histogram, we infer that average temperature of the simulation is indeed coming around the intended temperature viz. 600 K. Note that here, only the last 1 ps equilibrium stable data is considered after ignoring a large part of the data to avoid any thermal fluctuations. Therefore, Nose-Hoover thermostat is apt to control the temperature during simulation.

{}